\documentclass[pra,twocolumn,showpacs,showkeys,preprintnumbers,superscriptaddress]{revtex4}

\usepackage{graphicx}
\usepackage{bm}
\usepackage{amsmath}
\usepackage{amssymb}
\usepackage{float}
\usepackage[T1]{fontenc}

\begin{document}

\title{Entanglement Witnesses and Measures for Bright Squeezed Vacuum}

\author{Magdalena Stobi\'nska}
\affiliation{Institute of Physics, Polish Academy of Sciences, Al.\ Lotnik\'ow 32/46, 02-668 Warsaw, Poland}
\affiliation{Institute of Theoretical Physics and Astrophysics, University of Gda\'nsk, 80-952 Gda\'nsk, Poland}

\author{Falk T\"oppel}
\affiliation{Max Planck Institute for the Science of Light, G\"unther-Scharowsky-Stra\ss{}e 1/Bau 24, 91058 Erlangen, Germany}
\affiliation{University of Erlangen-N\"urnberg, Staudtstra\ss{}e 7/B2, 91058 Erlangen, Germany}

\author{Pavel Sekatski}
\affiliation{Group of Applied Physics, University of Geneva, Chemin de Pinchat 22, CH-1211 Geneva, Switzerland}

\author{Maria~V.~Chekhova}
\affiliation{Max Planck Institute for the Science of Light, G\"unther-Scharowsky-Stra\ss{}e 1/Bau 24, 91058 Erlangen, Germany}
\affiliation{Department of Physics, M.~V.~Lomonosov Moscow State University, Leninskie Gory, 119991 Moscow, Russia}

\begin{abstract}
Quantum entanglement is a fascinating phenomenon, especially if it is observed at the macroscopic scale. Importantly, macroscopic quantum correlations can be revealed only by accurate measurement outcomes and strategies. Here, we formulate feasible entanglement witnesses for bright squeezed vacuum in the form of the macroscopically populated polarization triplet Bell states. Their testing involves efficient photodetection and the measurement of  the Stokes operators variances. We also calculate the measures of entanglement for these states such as the Schmidt number and the logarithmic negativity. Our results show that the bright squeezed vacuum degree of polarization entanglement scales as the mean photon number squared. We analyze the applicability of an operational analog of the Schmidt number.
\end{abstract}


\maketitle

\section{Introduction}

Quantum entanglement is a fascinating phenomenon, especially if it is observed at the macroscopic scale~\cite{Schrodinger,Horodecki}. It emerges from the quantum superposition principle lying at the heart of quantum mechanics. If a two- or multimode superposition is highly populated, macroscopically entangled subsystems are created. A bipartite system maximally entangled in a given degree of freedom  has two subsystems, for which the values of the degree of freedom are completely random but still perfectly correlated. Importantly, macroscopic quantum correlations can be revealed only by accurate measurement outcomes and strategies~\cite{Ramanathan2011}.

Entanglement is the basic resource for quantum information processing, quantum communication and other quantum technologies~\cite{Gisin2002,Braunstein2005}. For this reason, there is an important need for efficient and reliable entanglement verification and quantification in various physical systems. This is done by the measurement of entanglement witnesses or measures~\cite{Horodecki}. While an entanglement witness only tells whether a system is entangled or not, an entanglement measure allows one to quantify the amount of entanglement. Unfortunately, most of the entanglement measures are not \textit{operational}, i.e.\ are not directly measurable in experiment. An operational measure of entanglement has been proposed for continuous variables such as wave vector or frequency~\cite{R}, but it is absent for discrete variables of large dimension, e.g.\ $10^{10}$, such as photon number.

The first condition for entanglement (inseparability), the Peres-Horodecki criterion, has been formulated in terms of density matrices~\cite{Peres&Horodecki} and therefore was not operational. Later, other conditions were formulated in terms of measurable quantities such as variances or uncertainties of ~\cite{Duan,R_Simon,Mancini,Agarwal}. A general method of finding inseparability criteria for bipartite continuous variable systems, deeply related to the Peres-Horodecki condition, was proposed in~\cite{Shchukin} and then amended in~\cite{MiranowiczComment,Miranowicz}. It contains a hierarchy of inequalities for the measurable moments of creation and annihilation operators and generalizes the previously obtained conditions~\cite{Duan,R_Simon,Mancini,Agarwal}.

Recently, macroscopic states of light generated a lot of interest and posed important questions in the scientific community. A possibility of verifying macroscopic entanglement became intriguing and was widely discussed~\cite{A,B,MDF}. Among macroscopic photonic states, two kinds of entangled states have been generated and reported: the micro-macro polarization singlet state, where a macroscopic qubit is entangled with a single photon~\cite{DeMartini2008}, and entangled bright squeezed vacuum (BSV) states, macroscopic analogs of two-photon polarization Bell states~\cite{Macrobell}. An entanglement test with the micro-macro singlet state was performed \cite{DeMartini2008}, but later refuted \cite{Sekatski2009}. It was shown that in the case of macroscopic states, inefficient detection may falsely reveal entanglement in separable states \cite{Pomarico2011} and that it is incapable of grasping their quantum character \cite{Raeisi2011}. New criteria for the test were formulated \cite{Sekatski2009,Spagnolo2010}, but not put in experiment. Entanglement in BSV states is known as twin-beam multiphoton entanglement~\cite{PanRMP} and manifests itself in perfect polarization correlations between two macroscopically populated beams. Macroscopic singlet polarization Bell state~\cite{Macrobell} is formally equivalent to the singlet state of a large spin, with the spin value being hugely uncertain. An entanglement witness in the form of spin inequality for BSV was discussed \cite{Reid2002,Simon,Gatti}. Recently, the non-separability witness derived in Ref.~\cite{Simon} was tested in experiment~\cite{Iskhakov_Archive} for this state. The form in which the non-separability witness was derived is inapplicable to the other three BSV states (macroscopic triplet polarization Bell states) experimentally obtained in Ref.~\cite{Macrobell}. Experimental realization of entanglement measures, e.g.\ the Schmidt number or other measures based on the eigenvalues of the density operator, for entangled macroscopic states of light remains an open issue.

In this paper, we formulate feasible non-separability (entanglement) witnesses for the macroscopic triplet polarization Bell states in the spirit of the Duan criterion. Their testing involves efficient photodetection and the measurement of the Stokes operators variances. We also calculate measures of entanglement for these states such as the logarithmic negativity and the Schmidt number as well as the operational measure $R$ introduced in Ref.~\cite{R}. Finally, we examine the dependence of the measure $R$ and its relation to the Schmidt number on the experimental conditions.

This paper is organized as follows. We start with introducing the basic properties of the bright squeezed vacuum in Section~\ref{sec1}. Section~\ref{sec2} is devoted to the derivation of the operational entanglement witnesses for the macroscopic triplet polarization Bell states. Next, in Section~\ref{sec3} we discuss various entanglement measures: the Schmidt number, the effective Schmidt number, and the negativity. A possibility of their experimental verification is addressed. Finally, we make the conclusions in Section~\ref{concl}.

\section{Entangled squeezed vacuum}\label{sec1}

Entangled four-mode BSV states of light are prepared in experiment by employing two two-mode BSV states~\cite{Braunstein} obtained via high-gain unseeded parametric down-conversion (PDC). In Ref.~\cite{Macrobell}, they were created by overlapping on a polarizing beam splitter two orthogonally polarized beams of frequency-nondegenerate squeezed vacuum.

The generation process, depending on the experimental conditions, is described by one of the Hamiltonians
\begin{align}
\mathcal{H}_{\Psi\pm} ={}& i \hbar g (a_H^{\dagger} b_V^{\dagger} \pm a_V^{\dagger} b_H^{\dagger}) + \mathrm{h.c.},
\label{PDC-Hamiltonian}
\\
\mathcal{H}_{\Phi\pm} ={}& i \hbar g (a_H^{\dagger} b_H^{\dagger} \pm a_V^{\dagger} b_V^{\dagger}) + \mathrm{h.c.},
\nonumber
\end{align}
where $g$ is the coupling constant proportional to the pump field, the PDC crystal length, and the second-order nonlinearity of the crystal. The down-converted photons carry linear polarization H (horizontal) and V (vertical) and are emitted in two frequency modes described by the creation operators $a^{\dagger}$ and $b^{\dagger}$. The resulting states can be considered as macroscopic (multiphoton) generalizations of the two-photon polarization Bell states,
\begin{align}
|\Psi^{(\pm)}_{mac}\rangle={}&e^{\Gamma(a_{H}^{\dagger} b_{V}^{\dagger}\pm
a_{V}^{\dagger}b_{H}^{\dagger})+\hbox{h.c.}}|\hbox{vac}\rangle,
\label{state}
\\
|\Phi^{(\pm)}_{mac}\rangle={}&e^{\Gamma(a_{H}^{\dagger} b_{H}^{\dagger}\pm
a_{V}^{\dagger}b_{V}^{\dagger})+\hbox{h.c.}}|\hbox{vac}\rangle,
\nonumber
\end{align}
where $\Gamma=\int g\,\mathrm{d}t$ is the parametric gain coefficient.

The above equation allows one to determine their Schmidt decomposition. For the singlet state $|\Psi_{mac}^{(-)}\rangle$, the decomposition is known~\cite{Iskhakov_Archive,Braunstein2005}. For $|\Psi_{mac}^{(+)}\rangle$, it has a similar form, so that both can be written as
\begin{equation}
|\Psi^{(\pm)}_{mac}\rangle=\sum_{n,m=0}^{\infty}(\pm1)^m\sqrt{\lambda_n\lambda_m}|n,m\rangle_a|m,n\rangle_b,
\label{Schmidt_s}
\end{equation}
where $\lambda_n\equiv\tanh^{2n}\Gamma/\cosh^2\Gamma$ and $|n,m\rangle_a\equiv|n\rangle_{a_H}\otimes|m\rangle_{a_V}$ denotes a two-mode Fock state  with $n$ photons polarized horizontally and $m$ photons polarized vertically in beam $a$ (similarly for beam $b$). It is possible to factorize Eq.~(\ref{Schmidt_s}) further into two independent Schmidt decompositions, one of them involving modes $a_H,b_V$ and the other one, modes $a_V,b_H$~\cite{Iskhakov_Archive},
\begin{align}
|\Psi^{(\pm)}_{mac}\rangle={}&|\Psi_1^{\pm}\rangle\otimes|\Psi_2^{\pm}\rangle,
\label{subsys}\\
|\Psi_1^{\pm}\rangle ={}& \sum_{n=0}^{\infty}\sqrt{\lambda_n}|n\rangle_{a_H}|n\rangle_{b_V},\nonumber\\
|\Psi_2^{\pm}\rangle ={}& \sum_{m=0}^{\infty}(\pm1)^m\sqrt{\lambda_m}|m\rangle_{a_V}|m\rangle_{b_H}.\nonumber
\end{align}
The Schmidt decompositions for the other two triplet states, $|\Phi^{(\pm)}_{mac}\rangle$, can be easily written by recalling that they are obtained from $|\Psi^{(+)}_{mac}\rangle$ by rotating the polarization. As a result, they will have the same form as the one for $|\Psi^{(+)}_{mac}\rangle$, but will be expressed in different polarization bases. For $|\Phi^{(+)}_{mac}\rangle$ it will be the right and left ($R,L$) circular polarization and for $|\Phi^{(-)}_{mac}\rangle$ it will be the $\pm45^{\circ}$ linear polarization basis.

The key difference between the two- and the four-mode BSV concerns entanglement. The two-mode BSV involves only photon-number entanglement. Namely, the photon number in the beam $a$ with a fixed polarization (e.g.\ $H$) is unknown, but it is always equal to the photon number in the beam $b$ having the orthogonal polarization ($V$). This state is known to approximate the maximally entangled EPR state in the high gain limit \cite{Englert2002}.
In the coordinate representation in the limit $\Gamma \to \infty$  the electric field quadratures become completely uncertain and $\delta$-correlated.
Four-mode BSV is a product of two such states and thus, it simply provides two copies of it. However, it also
contains polarization entanglement between beams $a$ and $b$. This polarization entanglement is probed through measuring the photon-number  correlations present in the two-mode BSVs (see Eq.~(\ref{Schmidt_s})). It is called twin-beam multiphoton entanglement~\cite{PanRMP} and is most easy to notice if e.g.\ $|\Psi^{(-)}_{mac}\rangle$ is re-written as a superposition
\begin{align}
|\Psi^{(-)}_{mac}\rangle ={}& \frac{1}{\cosh^2 \Gamma} \sum_{n=0}^{\infty} \sqrt{n+1} \tanh^{n} \Gamma |\psi^{(-)}_{n}\rangle,
\label{singlety}\\
|\psi^{(-)}_{n}\rangle ={}& \frac{1}{\sqrt{n+1}} \sum_{m=0}^n (-1)^m |n-m,m\rangle_a |m,n-m\rangle_b,
\nonumber
\end{align}
where $|\psi^{(-)}_{n}\rangle$ is an analog of a singlet state of two spin-$\frac{n}{2}$ particles. $|\Psi^{(-)}_{mac}\rangle$ is
invariant with respect to joint rotations of the polarization bases of both modes. Polarization of each beam separately is undetermined, but the polarizations of the beams $a$ and $b$ are anti-correlated. The situation is similar for the triplet states. This explains why these states can be considered as macroscopic (multiphoton) generalizations of the two-photon polarization Bell states.

\section{Entanglement witnesses}\label{sec2}

Entanglement witnesses are sufficient conditions for entanglement. Although sometimes being inconclusive, they are so far the only practical option for proving the entanglement of multidimensional systems. Conclusive (necessary and sufficient) conditions of entanglement are formulated only for special classes of quantum systems. For instance, the Peres-Horodecki criterion~\cite{Peres&Horodecki} provides a necessary and sufficient condition for the entanglement of two- and three-dimensional systems. Its continuous-variable counterpart, formulated by Simon~\cite{R_Simon}, is in the general case also only a sufficient condition, but becomes necessary for Gaussian states. In practice, continuous-variable entanglement is often witnessed using the Duan criterion~\cite{Duan}, containing variances of sum and difference quadratures for the subsystems of a bipartite system.

In experiment, the output states $|\Psi^{(\pm)}_{mac}\rangle$, $|\Phi^{(\pm)}_{mac}\rangle$ are given by a compound beam, comprising two independent frequency modes. Therefore, their Stokes operators $\mathcal{S}_i$ are given by the sum of \textit{partial}~\cite{Karassiov} Stokes operators for modes $a$ and $b$,
\begin{equation}
\mathcal{S}_i = \mathcal{S}_i^{a} + \mathcal{S}_i^{b},
\label{Stokes}
\end{equation}
with $i=0,1,2,3$, $\mathcal{S}_0^a = a_{H}^{\dagger}a_{H} + a_{V}^{\dagger}a_{V}$, $\mathcal{S}_1^a = a_{H}^{\dagger}a_{H} - a_{V}^{\dagger}a_{V}$, $\mathcal{S}_2^a = a_{H}^{\dagger}a_{V} + a_{V}^{\dagger}a_{H}$, $\mathcal{S}_3^a = i(a_{V}^{\dagger}a_{H} - a_{H}^{\dagger}a_{V})$, and similarly for mode b.

The condition involving variances of the Stokes operators,
\begin{equation}
\hbox{Var}(\mathcal{S}_1) + \hbox{Var}(\mathcal{S}_2) + \hbox{Var}(\mathcal{S}_3) \ge 2\mathcal{S}_0,
\label{var_in}
\end{equation}
holds true for any separable state of subsystems $a$ and $b$~\cite{Iskhakov_Archive,Simon}. This fact allows to formulate an entanglement witness operator
\begin{align}
\mathcal{W}_S ={}& (\mathcal{S}_1^a+\mathcal{S}_1^b-\langle\mathcal{S}_1^a+\mathcal{S}_1^b\rangle)^2 \nonumber\\
{}+{}& (\mathcal{S}_2^a+\mathcal{S}_2^b - \langle\mathcal{S}_2^a+\mathcal{S}_2^b\rangle)^2  \nonumber\\
{}+{}&(\mathcal{S}_3^a+\mathcal{S}_3^b-\langle\mathcal{S}_3^a+\mathcal{S}_3^b\rangle)^2-2\mathcal{S}_0.
\label{wit_sing}
\end{align}
As usual, negative mean value of the witness indicates entanglement.

It is worth noting that the sign of $\langle\mathcal{W}_S\rangle$ is invariant to the number of spatial and temporal modes because
the overall state is a product $\prod_k |\Psi^{(\pm)}_{k}\rangle$  of states for different modes that are pairwise entangled (between modes $a_{V,k}, a_{H,k}, b_{V,k}$ and $b_{H,k}$), so both the Stokes variances and the mean photon number of the whole beam contain additive contributions of separate modes. However, multimode separable states do not necessarily have this property. Therefore, the witness is valid for spatially and temporally multimode beams, under the assumption that separate modes are independent (the overall state is a product) \cite{Comment1,hidden}. This property enabled its experimental testing~\cite{Iskhakov_Archive} for the macroscopic BSV singlet state $|\Psi^{(-)}_{mac}\rangle$. Indeed, the macroscopic polarization singlet Bell state has completely suppressed noise in all Stokes observables $\mathcal{S}_{1,2,3}$~\cite{Macrobell}, hence $\langle\mathcal{W}_S\rangle = - 2\langle\mathcal{S}_0\rangle<0$.

At the same time, the witness~(\ref{wit_sing}) will not be negative for the three triplet states $|\Psi^{(+)}_{mac}\rangle$, $|\Phi^{(\pm)}_{mac}\rangle$, since they have noise suppressed only in one Stokes observable~\cite{hidden}. However, all states in Eq.~(\ref{state}) can be transformed into each other by local polarization transformations and thus, they contain the same amount of entanglement. Based on $\mathcal{W}_S$, we further derive the witnesses applicable to the triplet states.

 The $|\Psi^{(+)}_{mac}\rangle$ state and the singlet state are linked by a local unitary rotation  $|\Psi^{(+)}_{mac}\rangle = \mathbf{1}_a \otimes \mathcal{U}_b |\Psi^{(-)}_{mac}\rangle$, where $\mathcal{U}_b = e^{i \pi b_H^{\dagger}b_H}$, $\mathcal{U}_b^{\dagger} \mathcal{U}_b =\mathbf{1}_b$ and $\mathbf{1}_{a(b)}$ is a unity operator acting on beam $a$ ($b$). This follows from the fact that the rotation $\mathbf{1}_a \otimes \mathcal{U}_b$ transforms the Hamiltonian $\mathcal{H}_{\Psi-}$ into the Hamiltonian $\mathcal{H}_{\Psi+}$. In experiment, this transformation is easily realized by means of a half-wave plate inserted into beam $b$ with the optic axis oriented vertically or horizontally. Thus, the entanglement witness $\mathcal{W}_{T1}$ for  $|\Psi^{(+)}_{mac}\rangle$ equals $\mathcal{W}_{T1}=\mathcal{U}_b \mathcal{W}_S \mathcal{U}_b^{\dagger}$, which yields
\begin{align}
\mathcal{W}_{T1} ={}&(\mathcal{S}_1^a+\mathcal{S}_1^b-\langle\mathcal{S}_1^a+\mathcal{S}_1^b\rangle)^2 \nonumber\\
{}+{}& (\mathcal{S}_2^a-\mathcal{S}_2^b-\langle\mathcal{S}_2^a-\mathcal{S}_2^b\rangle)^2 \nonumber\\
{}+{}& (\mathcal{S}_3^a-\mathcal{S}_3^b-\langle\mathcal{S}_3^a-\mathcal{S}_3^b\rangle)^2 - 2\mathcal{S}_0.
\label{wit1}
\end{align}
Of course, in theory the measurement of the witness $\mathcal{W}_{T1}$ for the state $|\Psi^{(+)}_{mac}\rangle$ is equivalent to the measurement of the witness $\mathcal{W}_{S}$ for the singlet state, $\langle \Psi^-| \mathcal{W}_S |\Psi^-\rangle = \langle \Psi^+| \mathcal{W}_{T1} |\Psi^+\rangle$.

Entanglement witnesses for the other two macroscopic polarization triplet Bell states, $|\Phi^{(\pm)}_{mac}\rangle$,  can be easily obtained by recalling that the triplet states are transformed into each other via global rotations in the Stokes space. In particular, $|\Phi^{(+)}_{mac}\rangle$ is obtained from $|\Psi^{(+)}_{mac}\rangle$ by a $\pi/2$ rotation around the $\mathcal{S}_2$ axis~\cite{hidden}. This rotation can be realized by a quarter-wave plate with the optic axis at angle $45^{\circ}$ inserted into both beams $a,b$. The resulting witness can be obtained from~(\ref{wit1}) by changing variables $\mathcal{S}_1\rightarrow \mathcal{S}_3$ and $\mathcal{S}_3\rightarrow - \mathcal{S}_1$,
\begin{align}
\mathcal{W}_{T2} {}={}& (\mathcal{S}_1^a-\mathcal{S}_1^b-\langle\mathcal{S}_1^a-\mathcal{S}_1^b\rangle)^2 \nonumber\\
{}+{}& (\mathcal{S}_2^a-\mathcal{S}_2^b-\langle\mathcal{S}_2^a-\mathcal{S}_2^b\rangle)^2 \nonumber\\
{}+{}& (\mathcal{S}_3^a+\mathcal{S}_3^b-\langle\mathcal{S}_3^a+\mathcal{S}_3^b\rangle)^2 - 2\mathcal{S}_0.
\label{wit2}
\end{align}

Finally, $|\Phi^{(-)}_{mac}\rangle$ is obtained from $|\Psi^{(+)}_{mac}\rangle$ by a $\pi/2$ rotation around the $\mathcal{S}_3$ axis~\cite{hidden}. It can be realized by a $\pi/2$ rotator (a quartz crystal or a Faraday cell). The witness has the form
\begin{align}
\mathcal{W}_{T3} ={}& (\mathcal{S}_1^a-\mathcal{S}_1^b-\langle\mathcal{S}_1^a-\mathcal{S}_1^b\rangle)^2 \nonumber\\
{}+{}& (\mathcal{S}_2^a+\mathcal{S}_2^b-\langle\mathcal{S}_2^a+\mathcal{S}_2^b\rangle)^2 \nonumber\\
{}+{}& (\mathcal{S}_3^a-\mathcal{S}_3^b-\langle\mathcal{S}_3^a-\mathcal{S}_3^b\rangle)^2 - 2\mathcal{S}_0.
\label{wit3}
\end{align}

Indeed, the witnesses (\ref{wit1}-\ref{wit3}) have negative mean values for the corresponding triplet states, because $\hbox{Var}(\mathcal{S}_1^a+\mathcal{S}_1^b)=\hbox{Var}(\mathcal{S}_2^a-\mathcal{S}_2^b)=\hbox{Var}(\mathcal{S}_3^a-
\mathcal{S}_3^b)=0$ for state $|\Psi^{(+)}_{mac}\rangle$, $\hbox{Var}(\mathcal{S}_1^a-\mathcal{S}_1^b)=\hbox{Var}(\mathcal{S}_2^a-\mathcal{S}_2^b)=\hbox{Var}(\mathcal{S}_3^a+
\mathcal{S}_3^b)=0$ for state $|\Phi^{(+)}_{mac}\rangle$, and $\hbox{Var}(\mathcal{S}_1^a-\mathcal{S}_1^b)=\hbox{Var}(\mathcal{S}_2^a+\mathcal{S}_2^b)=\hbox{Var}(\mathcal{S}_3^a-
\mathcal{S}_3^b)=0$ for state $|\Phi^{(-)}_{mac}\rangle$.

It is worth mentioning that the conditions for entanglement given by witnesses (\ref{wit_sing}-\ref{wit3}) look very similar to the Duan condition as they contain the variances of sum and difference operators for subsystems $A$ and $B$. At the same time, the relation between the Stokes operators from (\ref{wit_sing}-\ref{wit3}) and the quadrature operators from the Duan condition is only known for the case of light with a bright polarized coherent component and not for our case.

\begin{figure}[h]
\includegraphics[width=0.35\textwidth]{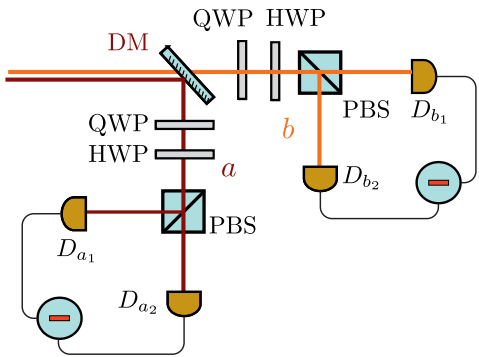}
\caption{(Color online) Experimental setup for testing the entanglement witnesses~(\ref{wit1}-\ref{wit3}). In each beam, there is a Stokes measurement setup: PBS, polarizing beam splitter; HWP, half-wave plate; QWP, quarter-wave plate; D, detector. The signals from the two detectors in each arm are subtracted to obtain the Stokes observables.} \label{setup}
\end{figure}

In Fig.~\ref{setup} we depicted an experimental setup where the witnesses $\mathcal{W}_{T}$ can be tested. It enables simultaneous measurement of
various Stokes observables for beams $a$ and $b$. The complete test will consist of the measurement of the variances for combinations of partial Stokes operators, $\hbox{Var}(\mathcal{S}_i^a\pm\mathcal{S}_i^b)$, $i=1,2,3$, and the total photon number $\langle\mathcal{S}_0\rangle$.
The measurement should consist of three series. In each series, one of the variances entering Eqs.~(\ref{wit1}-\ref{wit3}) is measured. This implies certain positions of the wave plates in the Stokes measurement setup. For $\mathcal{S}_1$ measurement, both wave plates should have their optic axes horizontal. For $\mathcal{S}_2$ measurement, the HWP should have the optic axis at $22.5^{\circ}$ and the QWP, at $45^{\circ}$. Finally, for $\mathcal{S}_3$ measurement, the HWP should have the optic axis horizontal and the QWP, at $45^{\circ}$. The variances should be calculated by averaging over a large number of pulses.

\section{Entanglement measures} \label{sec3}

\subsection{Schmidt number}

One of the well-known entanglement measures is the number of nonzero Schmidt coefficients $\sqrt{\lambda_i}$ in the Schmidt decomposition~\cite{Horodecki}. It is called the Schmidt rank or number. In case of a maximally entangled bipartite system with symmetrical subsystems, defined in the Hilbert space $\mathcal{H}=\mathcal{C}^d \otimes \mathcal{C}^d$, all Schmidt coefficients have to be equal and the Schmidt number is $K=d$, where $d$ is the dimensionality of the subsystem. In general, $1\le K \le d$ and a state is separable if $K=1$. To quantify entanglement in systems with infinite dimensional Hilbert space such as BSV, for which $K=\infty$, another measure is more appropriate, which we will further call the \textit{effective Schmidt number}. It is defined as follows~\cite{Rzazewski,Law}
\begin{equation}
\bar{K} \equiv 1/\mathrm{Tr}(\rho^2) = 1/\sum_{i} \lambda_i^2,
\end{equation}
where $\rho$ is a density operator and $\sum_{i} \lambda_i =1$. This definition coincides with the original definition of $K$ in the following way. For a separable state $\bar{K}=1$. For a maximally entangled state with all Schmidt coefficients equal and $d \to \infty$ (such a state does not exist because it is not normalizable) we would obtain $\bar{K}=d \to \infty$. Otherwise, $\bar{K}$ is finite even if the number of Schmidt coefficients $\sqrt{\lambda_i}$ is infinite.

The effective Schmidt number for $|\Psi^{(\pm)}_{mac}\rangle$ state is the product of the effective Schmidt numbers for subsystems $|\Psi_{1,2}^{\pm}\rangle$, $\bar{K}_1=\bar{K}_2=[\sum_{n=0}^{\infty}\lambda_n^2]^{-1}$, which yields $\bar{K}=(1+2\sinh^2\Gamma)^2$. Since all the macroscopic polarization Bell states~(\ref{state}) have the same form of the Schmidt decomposition, the effective Schmidt number for all of them is the same, $\bar{K}=(1+2N_0)^2$, where $N_0=\sinh^2\Gamma$ is the photon population in each mode $a_H,a_V,b_H,b_V$. For bright states $N_0\gg1$ and $\bar{K}\approx4N_0^2$, hence the degree of polarization entanglement grows quadratically with the mean photon number.

\subsection{Negativity}

Other widely used and easily calculated measures of entanglement are the negativity and the logarithmic negativity which gives an upper bound for distillable entanglement~\cite{vidal_werner}. For a bipartite quantum state $\rho$ they are defined as $\mathcal{N}(\rho)=\|\rho^{PT}\|_1-1$ and $E_\mathcal{N}(\rho)=\log_2\|\rho^{PT}\|_1$, where $PT$ denotes partial transposition with respect to one of the subsystems and $\|\mathcal{A}\|_1=\text{Tr}\sqrt{\mathcal{A}^\dagger\mathcal{A}}$ is the trace norm of an operator $\mathcal{A}$. Particularly useful is the fact that the trace norm is factorizable, $\|\rho_1\otimes \rho_2\|_1= \|\rho_1\|_1 \|\rho_2\|_1$, and that the logarithmic negativity is additive, i.e.,\ $E_\mathcal{N}(\rho_1\otimes \rho_2)=E_\mathcal{N}(\rho_1)+E_\mathcal{N}(\rho_2)$~\cite{vidal_werner}.

Since four-mode BSV is a product of two entangled bipartite subsystems as in~(\ref{subsys}), it is sufficient to determine the negativities of each of them, $\rho_{1,2}$, separately. Performing partial transpositions with respect to the subsystems $b_V$ and $b_H$, respectively, we obtain $\|\rho_1^{PT}\|_1=\|\rho_2^{PT}\|_1=\left[\sum_{n=0}^\infty\sqrt{\lambda_n}\right]^2=\text{e}^{2\Gamma}$. Thus, for the four-mode BSV, the
negativity equals $\mathcal{N}(\rho)= \text{e}^{4\Gamma}-1$. For high gain, $\mathcal{N}(\rho) \simeq 16 \sinh^4 \Gamma = 16 N_0^2$. Again, the quadratic dependence of the degree of entanglement on the population is observed. We notice that $\mathcal{N}(\rho) = 4 \bar{K}$. The logarithmic negativity takes the value $E_\mathcal{N}(\rho)=4 \Gamma/\ln 2$. In comparison, for two-mode BSV it equals $2\Gamma/\ln 2$~\cite{prauzner_bechcicki}, which shows that BSV macroscopic Bell states contain (with respect to the logarithmic negativity) twice more entanglement than the usual two-mode BSV.
This is understandable: since four-mode BSV consists of two copies of entangled states, twice more entanglement can be distilled from it than from a single copy.

\subsection{Fedorov ratio}

Both the Schmidt number and the negativity are not operational quantifiers of entanglement as they cannot be directly measured in experiment. For a bipartite system entangled in a continuous variable, an operational measure was proposed~\cite{R}, called the Fedorov ratio. It is defined in the spirit of the entropy of entanglement with the advantage of being directly measurable in experiment. Further, we try to adapt this measure for characterizing BSV, entangled in the photon number, a variable that is discrete but can be viewed as pseudo-continuous when large numbers are involved.

Consider a pure bipartite quantum system, entangled in a continuous variable $\nu$, with the state vector
\begin{equation}
|\Psi\rangle=\int d\nu_a d\nu_b F(\nu_a,\nu_b)|\nu_a\rangle|\nu_b\rangle.
 \label{nu}
\end{equation}
The variable $\nu_a$ can be characterized by its marginal probability distribution $P(\nu_a)=\int d\nu_b|F(\nu_a,\nu_b)|^2$, with the standard deviation $\Delta\nu_a$, and the conditional probability distribution $P(\nu_a|\nu_b)=|F(\nu_a,\nu_b)|^2$, evaluated for a certain value of $\nu_b$, with the width $\delta\nu_a$.

The effective Schmidt number $\bar{K}$ for the state $|\Psi\rangle$ is very well approximated by the ratio $R$~\cite{R} defined as
\begin{equation}
R_{\nu}=\frac{\Delta\nu_a}{\delta\nu_a}.
\label{R}
\end{equation}
Of course, equivalently the variable $\nu_b$ may be involved in this definition, instead of $\nu_a$. The parameter $R$, known as the Fedorov ratio, can be easily obtained in experiment and is hence an \textit{operational measure of entanglement}. Note however, that this measure is only defined for pure states, so to be operational stricto sensu it has to be supplemented with an experimental proof of the purity of the global state. The possibility to generalize the Fedorov ratio to mixed states is an open question. For instance, it has been measured for the cases of wave vector~\cite{Straupe} and frequency entanglement~\cite{Brida}. For Gaussian states, the Fedorov ratio exactly coincides with the effective Schmidt number~\cite{Gauss}.

We adopt the definition~(\ref{R}) for the photon-number variable and its probability distributions in the following way:
\begin{equation}
R_n=\frac{\Delta n_a}{\delta n_a},
 \label{R_n}
\end{equation}
where $\Delta n_a$ is the width of the marginal photon-number distribution in beam $a$, while $\delta n_a$ is the width of the photon-number distribution in beam $a$ under the condition that a certain photon number $n_b$ has been measured in beam $b$.

\begin{figure}[h]
\includegraphics[width=0.4\textwidth]{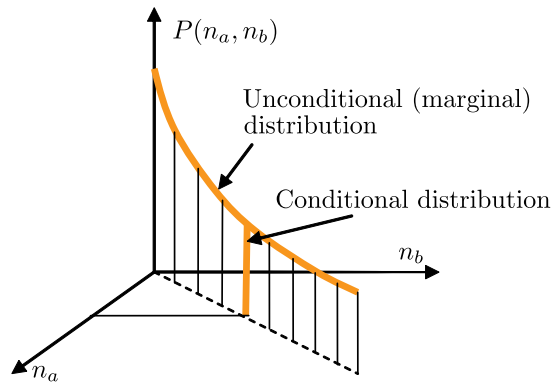}
\caption{(Color online) Photon-number distributions for the states $|\Psi_{1,2}^{\pm}\rangle$.}
\label{distributions}
\end{figure}

Similarly to the effective Schmidt number, the ratio $R_n$ for each of the macroscopic Bell states is a product of the ratios $R_{n1}$ and $R_{n2}$ of its two subsystems. For instance, for the state $|\Psi^{(\pm)}_{mac}\rangle$, the states $|\Psi_{1,2}^{\pm}\rangle$ in (\ref{subsys}) have the same Fedorov ratios. They can be easily calculated by noticing that the marginal distribution $P(n_a)$ is a geometric one,
\begin{equation}
P(n_a)=\frac{(\tanh\Gamma)^{2n_a}}{\cosh^2\Gamma},
 \label{marginal}
\end{equation}
while the conditional distribution is given by the Kronecker delta,
\begin{equation}
P(n_a|n_b)=\delta_{n_a, n_b}.
 \label{conditional}
\end{equation}
These distributions are schematically shown in Fig.~\ref{distributions}.

Assuming that the width of the discrete $P(n_a|n_b)$ distribution is unity, the ratios $R_{n1,2}$ are given by the standard deviation of the geometric distribution~(\ref{marginal}), $R_{n1,2}=\sqrt{2} \sinh^2\Gamma$. Finally, the Fedorov ratio for $|\Psi^{(\pm)}_{mac}\rangle$ equals
\begin{equation}
R_n=2 N_0^2.
 \label{ratio}
\end{equation}
Thus, in case of high population, the operational measure $R_n$ differs from the effective Schmidt number $\bar{K}$ by only a constant factor $1/2$.

The $R_n$ ratio for entangled BSV can be measured with the help of a setup shown in Fig.~\ref{setup}. The orientation of the HWP and QWP should be such that proper polarization bases are chosen in the arms $a$, $b$. For instance, in the case of the singlet state $|\Psi^{(-)}_{mac}\rangle$, the plates can be oriented in any way but similarly for arms $a$, $b$. The ratios $R_{n1}$ and $R_{n2}$ can then be measured independently using pairs of detectors $D_{a1},D_{b2}$ and $D_{a2},D_{b1}$. For each pair, after acquiring a certain (large) number of pulses, the photon-number distributions should be analyzed and the conditional and unconditional widths should be measured. In practice, because the detectors do not distinguish between close photon numbers, the photon-number distribution should be binned in the intervals of about $200$ photons~\cite{Iskhakov_Archive}.

\begin{figure}[h]
\includegraphics[width=0.4\textwidth]{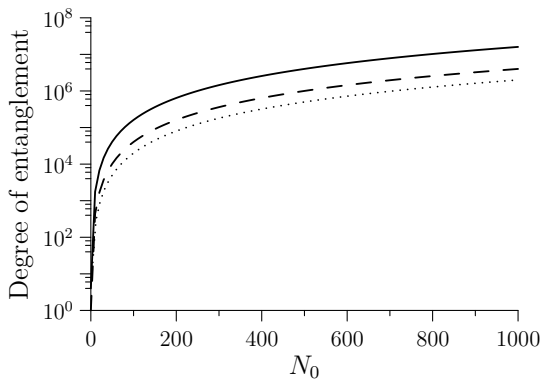}
\caption{The degree of polarization entanglement in the logarithmic scale for the four-mode entangled BSV state as a function of the average photon number $N_0$:  the negativity $\mathcal{N}$ (solid line), the effective Schmidt number $\bar{K}$ (dashed line) and the Fedorov ratio $R_n$ (dotted line).}
\label{compare}
\end{figure}

All the above-considered measures of entanglement for the case of $|\Psi^{(\pm)}_{mac}\rangle$ states are plotted in Fig.~\ref{compare} as functions of the mean photon number.

\subsection{Effective dimensionality of BSV Hilbert space}

Entangled BSV states of light are considered to be macroscopic generalizations of polarization singlet or triplet Bell states due to the symmetry reasons. Their polarization correlations are probed through photon-number measurements in polarization modes. Since the Hilbert space of these states is complex~\cite{complex} and infinite, it is interesting to understand and quantify the amount of their accessible entanglement for example, by comparison with finite-dimensional systems, where the notion of a maximally entangled state is well understood.

We propose a rough estimate for the dimensionality $d$ of the Hilbert space of $|\Psi^{(\pm)}_{mac}\rangle$, depending on the gain $\Gamma$, based on the following argument. In two-photon experiments, small gain $\Gamma \lesssim 10^{-3}$ is used to produce a superposition of the vacuum (the dominant component) and a biphoton. All higher-order contributions are largely suppressed. Truncation of the Fock states of order higher than one in Eq.~(\ref{Schmidt_s}), for a given gain, is justified if the normalization of the truncated state is preserved to a good approximation, given by a parameter~$\epsilon$
\begin{equation}
\lambda_0^2 + 2\lambda_0 \lambda_1 = 1 - \epsilon.
\end{equation}
For the two-photon case $\epsilon \approx \Gamma^4$. However, it is known that in these experiments the four- and six-photon components are observed as well, which is manifested as, e.g., a decrease in the interference visibility for relatively bright PDC sources~\cite{Laskowski2009}. Thus, in our example, the value of $\epsilon$ shows, for a given gain, how well the outcoming state from the PDC crystal can be approximated by a superposition of a two-photon state and the vacuum. Of course, the smaller $\epsilon$ is, the better the approximation gets.

By analogy with the low-gain case, for any value of $\Gamma$ we can locally restrict the state (\ref{Schmidt_s}) to some finite-dimensional Hilbert space $\mathcal{H}=\mathcal{C}^{d_a} \otimes \mathcal{C}^{d_b}$, where $d_a$ and $d_b$ denote dimensionalities of beams $a$ and $b$. Then, $1-\epsilon$ gives the probability to find the state in $\mathcal{H}$. The natural choice for the subspace $\mathcal{C}^{d_a}$ is to keep in $|\Psi^{(\pm)}_{mac}\rangle$ only these components which have a limited number of photons in beam $a$, $a_H^\dag a_H + a_V^\dag a_V \leq N_{max}$. Similarly for $\mathcal{C}^{d_b}$ and beam $b$. This restriction implies the following normalization condition for the truncated state $|\Psi^{T(\pm)}_{mac}\rangle$:
\begin{equation}
\sum_{n=0}^{N_{max}} \lambda_n \sum_{m=0}^{N_{max} - n} \lambda_m = 1 - \epsilon.
\label{norm}
\end{equation}
The normalization is calculated over the sectors of the density matrix with fixed photon number, so that $n+m \le N_{max}$. It enables one to determine the dimensionality of $\mathcal{C}^{d_{a,b}}$, $d=d_a = d_b = \frac{1}{2}(N_{max}+1)(N_{max}+2)$, and the dependence of $N_{max}$ on the average population $N_0$. Using $\tanh^2 \Gamma = N_0/(N_0+1)$ we turn Eq.~(\ref{norm}) into
\begin{equation}
\epsilon =\left(\tfrac{N_0}{N_0+1}\right)^{1+N_{max}\!\!}\!\left(N_{max}\!\!+2- \tfrac{N_0}{N_0+1}(N_{max}\!\!+1)\right)
\end{equation}
and obtain a linear relation between $N_{max}$ and $N_0$. It allows one to express the dimensionality for large population as $d~\approx~\frac{\alpha^2(\epsilon)}{2}N_0^2$, where $\alpha$ is a function of $\epsilon$ given by the equation $\epsilon = e^{-\alpha}(\alpha+1)$. For example, if $\epsilon=10^{-12}$ as is for the two-photon case, $\alpha \approx 31$. If $\epsilon=10^{-2}(10^{-1})$, we obtain $\alpha \approx 7 (4)$. Please note the quadratic scaling of the dimensionality with $N_0$.

The above estimations of the effective dimensionality of the Hilbert space for BSV are useful for reconstructing its density matrix, or the most significant part of it. It would provide almost full information about the joint photon-number distribution and could enable calculation of the entanglement measures based on the eigenvalues of the density operator obtained from the experimental data.

\begin{figure}[t]
\includegraphics[width=0.35\textwidth]{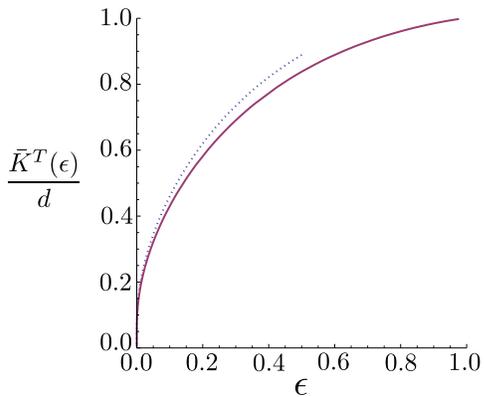}
\caption{(Color online) $\bar{K}^T/ d$ as a function of $\epsilon$. The solid line is for $N_0=10$, the dotted line $N_0=1$, for large $N_0$ ($>5$) the curve does not change with $N_0$.}
\label{maximality}
\end{figure}

Now, we investigate the amount of entanglement in the truncated state
\begin{equation}
|\Psi^{T(\pm)}_{mac}\rangle = \tfrac{1}{\sqrt{1-\epsilon}}\kern-1em\sum_{n,m=0}^{n+m \leq N_{max}}\kern-1.5em(\pm 1)^m \sqrt{\lambda_n \lambda_m}
|n,m\rangle_a|m,n\rangle_b.
\end{equation}
Since the restriction is local, the amount of entanglement in $|\Psi^{T(\pm)}_{mac}\rangle$ gives a lower bound on the overall entanglement in $|\Psi^{(\pm)}_{mac}\rangle$. The effective Schmidt number for the truncated state $\bar{K}^T = (1-\epsilon)^2\left(\sum_{n,m=0}^{n+m \leq N_{max}}  \lambda_n^2 \lambda_m^2 \right)^{-1}$ fulfills $(\tfrac{1-\epsilon}{1+\epsilon})^2 \bar{K} \le \bar{K}^T < (1-\epsilon) \bar{K}$.
We look at how far is the restricted state $|\Psi^{T(\pm)}_{mac}\rangle$ from a maximally entangled state in $\mathcal{H}$ as a function of $\epsilon$, where $\mathcal{H} = \mathcal{C}^{d} \otimes \mathcal{C}^{d}$. This question may be understood in a more practical way as showing how good is our approach for creating maximally entangled state of dimension $d$. Since the maximally entangled states are defined with respect to  dimensionality of their Hilbert space, we depicted $\bar{K}^T/ d$ as a function of $\epsilon$ in Fig.~\ref{maximality}. We notice that at high gain, a perfect maximally entangled state ($\bar{K}^T/ d=1$) is obtained only with $\epsilon$ approaching one. This is because $|\Psi^{T(\pm)}_{mac}\rangle$ is a superposition of singlet states of different spin values and thus belonging to Hilbert spaces with different dimensionalities. Moreover, the genuine maximally entangled d-dimensional singlet $|\psi^{(-)}_{N_{max}}\rangle$ from Eq.~(\ref{singlety}) has the smallest weight in the superposition. Thus, the state $|\Psi^{T(\pm)}_{mac}\rangle$ will certainly appear to be maximally entangled in detection, where it is projected on a fixed photon number subspace, but it is non-maximally entangled as a whole. An interesting fact is that the tradeoff between $\epsilon$ and $\bar{K}^T/ d$ is independent of $N_0$ above a certain value, which suggests that the ``maximality'' of entanglement in the discrete Hilbert space sense is unchanged when the gain increases.

\section{Conclusions}\label{concl}

In this paper we have discussed feasible entanglement witnesses and measures for bright squeezed vacuum in the form of macroscopic polarization Bell states, easy to generate in laboratories by means of parametric down conversion. Up to date, this is the only way to experimentally produce macroscopic entanglement (quantum correlations between two macroscopically populated subsystems) for light. In general, this kind of entanglement is known to be very difficult to produce and to verify, since the more macroscopic it is, the more it is fragile to disturbances and the more it requires measurement techniques with very high precision. The last condition is difficult to fulfill with current-technology photodetection.

We derived the entanglement witnesses for the triplet BSV states and suggested their implementation based on the measurement of Stokes operators variances. This detection method is very convenient for intense fields because it allows one to eliminate high intensity fluctuations present in other techniques, e.g.\ in direct intensity measurements. We also discussed the entanglement measure in the form of the effective Schmidt number, and found that it is very close to the Fedorov ratio, based on the measurement of conditional and unconditional photon-number probability distributions. We conclude that it can be considered to be an operational counterpart of the Schmidt number. Our results show that for all used measures, the BSV degree of entanglement scales as the mean photon number squared.

We hope that the presented ideas will open the way for efficient entanglement verification for other infinite dimensional systems.

\acknowledgments

This work was supported by the EU 7FP Marie Curie Career Integration Grant No. 322150 "QCAT", MNiSW grant No. 2012/04/M/ST2/00789 and FNP Homing Plus project. M.C. acknowledges the support by RFBR grant No.\ 12-02-00965-a.


\begin{thebibliography}{99}
\bibitem{Comment1} This assumption is indeed necessary: Consider the following multimode state with two photons on each side:
\begin{equation}
\frac{1}{2}(a_{H,1}^\dag a_{V,2}^\dag -a_{V,1}^\dag a_{H,2}^\dag)\lvert 0\rangle(b_{H,1}^\dag b_{V,2}^\dag -b_{V,1}^\dag b_{H,2}^\dag)\lvert 0\rangle.
\nonumber
\end{equation}
For this state the witness (\ref{wit_sing}) takes a negative value $\langle W_S \rangle~\!=~\!-8$, while the subsystems $A$ and $B$ are separable. In the multimode case the Stokes operators in (\ref{Stokes}) are given by the sum of Stokes operators for all different subsystems ($a_k$ and $b_k$), and the violation of the witness only ensures that there is entanglement between some subsystems.
\bibitem{complex} We mean that the states are defined in the photon-number spaces of four modes, $a_H, a_V, b_H, b_V$.
\end{thebibliography}
\end{document}